\documentclass{emulateapj}

\usepackage{amsmath}

\newcommand{\Mej}{$M_{\mathrm{ej}}$}
\newcommand{\Ekin}{$E_{\mathrm{kin}}$}
\newcommand{\Msun}{$M_{\odot}$}
\newcommand{\Mms}{$M_{\mathrm{MS}}$}
\newcommand{\Mni}{$M_{\mathrm{{}^{56}Ni}}$}
\newcommand{\Ni}{$^{56}$Ni}

\def\ion#1#2{{\rm #1}~{\sc #2}}

\shorttitle{SN 2007bi}
\shortauthors{Moriya et al.}

\begin{document}

\title{A Core-Collapse Supernova Model for the Extremely Luminous Type
  Ic Supernova 2007bi: An Alternative to the Pair-Instability Supernova Model}

\def\ipmu{1}
\def\ut{2}
\def\resceu{3}
\def\konan{4}

\author{
{Takashi Moriya}\altaffilmark{\ipmu,\ut,\resceu}, 
{Nozomu Tominaga}\altaffilmark{\konan,\ipmu},
{Masaomi Tanaka}\altaffilmark{\ipmu},
{Keiichi Maeda}\altaffilmark{\ipmu}, and
{Ken'ichi Nomoto}\altaffilmark{\ipmu,\ut}
}

\altaffiltext{\ipmu}{
Institute for the Physics and Mathematics of the Universe, University of
Tokyo, Kashiwanoha 5-1-5, Kashiwa, Chiba 277-8583, Japan;
takashi.moriya@ipmu.jp
}
\altaffiltext{\ut}{
Department of Astronomy, Graduate School of Science, University of Tokyo,
Bunkyo-ku, Tokyo 113-0033, Japan
}
\altaffiltext{\resceu}{
Recearch Center for the Early Universe,
Graduate School of Science, University of Tokyo,
Bunkyo-ku, Tokyo 113-0033, Japan
}
\altaffiltext{\konan}{
Department of Physics, Faculty of Science and Engineering, Konan University, 8-9-1 Okamoto, Kobe, Hyogo 658-8501, Japan
}

\begin{abstract}

We present a core-collapse supernova model for the extremely luminous
Type Ic supernova 2007bi.  By performing numerical calculations of
hydrodynamics, nucleosynthesis, and radiation transport, 
we find that SN 2007bi is consistent
with the core-collapse supernova explosion of a 43 \Msun~
carbon and oxygen core obtained from the
evolution of a
progenitor star with a main sequence mass of 100 \Msun~and metallicity of
$Z=Z_\odot/200$, from which its hydrogen and helium envelopes
are artificially stripped.
The ejecta mass and
the ejecta kinetic energy of the models are 40~\Msun~and $3.6\times
10^{52}$ erg.  The ejected \Ni~mass is as large as ~6.1~\Msun,
which results from the explosive nucleosynthesis with large explosion
energy. We also confirm that SN
2007bi is consistent with a pair-instability supernova model as has
recently been claimed.  We show that the earlier light curve data can
discriminate between the models for such luminous supernovae.

\end{abstract}

\keywords{supernovae: general, supernovae: individual (SN 2007bi, SN
2006gy), gamma-ray burst: general
}

\section{Introduction}
A massive star with the main sequence mass (\Mms) in
the range of 10 $-$ 140 \Msun~
forms an Fe core in its center and eventually collapses.
This collapse is thought to end up with
the core-collapse supernova (SN) of Type II, Ib, or Ic
(Filippenko 1997 for a review).
If a star is as massive as \Mms $= 140 - 300$ \Msun,
the oxygen-rich core becomes dynamically unstable owing to the 
electron-positron pair creations
(Rakavy \& Shaviv 1967; Barkat, Rakavy, \& Sack 1967).
As the internal energy is spent by the pair creations,
the core loses the stability and starts to collapse.
When the central temperature exceeds $\sim 5\times 10^{9}$ K,
the core becomes stable but the temperature
is so high that oxygen burning becomes explosive
to produce enough energy to unbind the star.
A large amount of \Ni~is synthesized by
the explosive burning
({\it e.g.}, Umeda \& Nomoto 2002, hereafter UN02;
Heger \& Woosley 2002) and the subsequent radioactive decays power the light curve.
Thus this event is theoretically predicted to be observed
as a pair-instability supernova (PISN).
Some luminous SNe like SN 2006gy
(Ofek et al. 2007; Smith et al. 2007)
have been suggested to be PISNe (see Section 4.2), but no clear consensus
has been reached ({\it e.g.}, Kawabata et al. 2009).

Recently, Gal-Yam et al. (2009) (G09 hereafter)
suggested that the extremely luminous Type Ic SN 2007bi is the first observed example of the PISN.
They showed that the PISN model is consistent
with the observed light curve (LC) and the nebular spectra of SN 2007bi.
They estimated the masses of C, O, Na, Mg, Ca, and \Ni~
from the observed optical spectra.
Other elements with no strong emission lines in the optical range,
Si and S, are assumed to be the same as
the PISN model of Heger \& Woosley (2002).
Young et al. (2010) (Y10 hereafter) showed multi-color observations of
SN 2007bi and the metallicity of the host galaxy.
However, the above observations of SN 2007bi have not
quantitatively been compared with the core-collapse SN models. In view
of the importance of clarifying the final fates of very massive stars,
we examine how strong the observational constraint on the theoretical
models are.

The aim of this Letter is to show that a core-collapse SN model is
indeed consistent with the observed properties of SN 2007bi, if the
progenitor is as massive as \Mms $\sim 100$ \Msun~ and the explosion
energy is large. This would imply that SN 2007bi might not
necessarily be a PISN.
In Section 2, we summarize the progenitor models for SN 2007bi and
numerical methods used for our calculations of hydrodynamics,
nucleosynthesis, and the LC.  The core-collapse SN models of
SN 2007bi are presented in Section 3 and the results are discussed in
Section 4.

\section{Progenitor and Explosion Modeling}
\subsection{Progenitor}
The high peak luminosity and the long rise time of the LC
of SN 2007bi (G09, Y10) require a large amount of \Ni~($>3$ \Msun, G09) and
a large ejecta mass.
These observations imply that the progenitor of SN 2007bi
is massive.
We apply a pre-SN model with
\Mms~=~100~\Msun~
calculated by Umeda \& Nomoto (2008, UN08 hereafter).
UN08 assumed the metallicity of the progenitor models to be
$Z=Z_\odot/200$, which is small enough to avoid a large amount of wind
mass loss.
Then the pre-SN model remains as massive as $M = 83$ \Msun,
whose carbon + oxygen (C+O) core is massive enough (43 \Msun) to produce
a large amount of \Ni.

However, the pre-SN model has a massive H-rich envelope, while
SN 2007bi does not show the lines of either H or He.
Therefore the progenitor must have lost its
H-rich envelope (36 \Msun) and He layer (4 \Msun) during
the pre-SN evolution, thus having only the bare C+O core
at the explosion.  We construct the pre-SN C+O star model of 43 \Msun,
by removing the H-rich envelope and He layer from the 83 \Msun~ star.
Note that the metallicity of the host galaxy of SN 2007bi
($Z\sim Z_\odot/3$, Y10) is higher than that of our
adopted progenitor ($Z=Z_\odot/200$).
The wind mass loss is expected to work more efficiently
and the main sequence mass of the progenitor which
has the C+O core mass of 43 \Msun~might be more massive.
The rotation of stars can also play a role
in the mass loss ({\it e.g.}, Meynet et al. 2003; Hirschi et al. 2004;
Maeder et al. 2005; Georgy et al. 2009).
Another possible cause of such envelope stripping is the formation of a
common envelope during a close binary system, where the smaller mass
companion star spirals into the envelope of the more massive star.
The outcome depends on whether the energy available from the spiral-in
exceeds the binding energy of the common envelope, thus being either a
merging of the two stars or the formation of two compact stars, {\it e.g.},
a C+O star and a He star.

\begin{figure}[t]
\begin{center}
\epsscale{1.}
\plotone{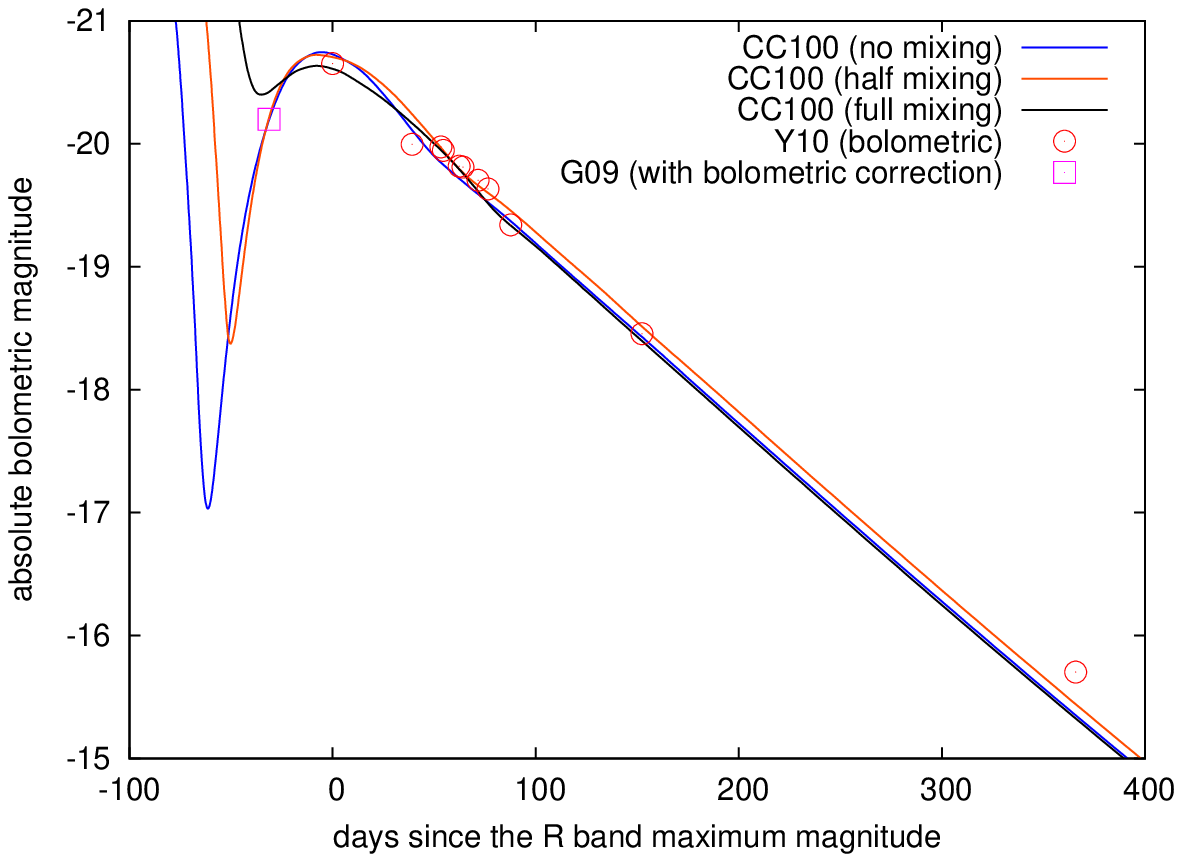}
\caption{
Bolometric LCs of the C+O star SN models CC100
(\Mej~=~40~\Msun, \Ekin~=~$3.6\times 10^{52}$ erg, and \Mni~=~6.1~\Msun).
The observed bolometric LC (open circles) is taken from Y10.
The bolometric magnitude of the rise part of SN 2007bi (open square)
is estimated from the $R$ band magnitude.
All the calculated LC has the same physical structure but
the degrees of mixing are different.
The horizontal axis shows the days in the rest frame.
}\label{07bi-ccLC}
\end{center}
\end{figure}

\subsection{Explosion Modeling}
We calculate the explosion of the pre-SN C+O star (43 \Msun)
as described above.
Explosions are induced by a thermal bomb and followed by a
one-dimensional Lagrangian code with the piecewise parabolic method
(Colella \& Woodward 1984).
Note that the explosion energy is a free parameter in core-collapse SN
explosion models while it is not in PISN explosion models.
Explosive nucleosynthesis is calculated as post-processing
for the thermodynamical history obtained by
the hydrodynamical calculations.
The resultant abundance distribution is basically very similar 
to those calculated by UN08 (see Figures 5 and 6 in UN08).

The dynamics of the ejecta is followed
until 1 day after the explosion, when the expansion already becomes
homologous ($r\propto v$).
The bolometric LCs are calculated for the homologous ejecta by using
the LTE radiation transfer code (Iwamato et al. 2000) that includes
the radioactive decays of \Ni~and $^{56}$Co as energy sources.  This
code calculates the $\gamma$-ray transport for a constant $\gamma$-ray
opacity (0.027 $\mathrm{cm^{2}~g^{-1}}$, Axelrod 1980) and assumes
all the emitted positrons are absorbed {\it in situ}\footnote
{This assumption of the positron absorption does not have much effect
on the LCs we show in this paper, because the contribution from
the gamma-rays is still a dominant energy source of them.
}
.  For the optical
radiation transport, the Thomson scattering opacity is obtained by
calculating the electron density from the Saha equation, and the
Rosseland mean opacity is estimated from the empirical relation to the
Thomson scattering opacity (Deng et al. 2005).

\begin{figure}[t]
\begin{center}
\epsscale{1.}
\plotone{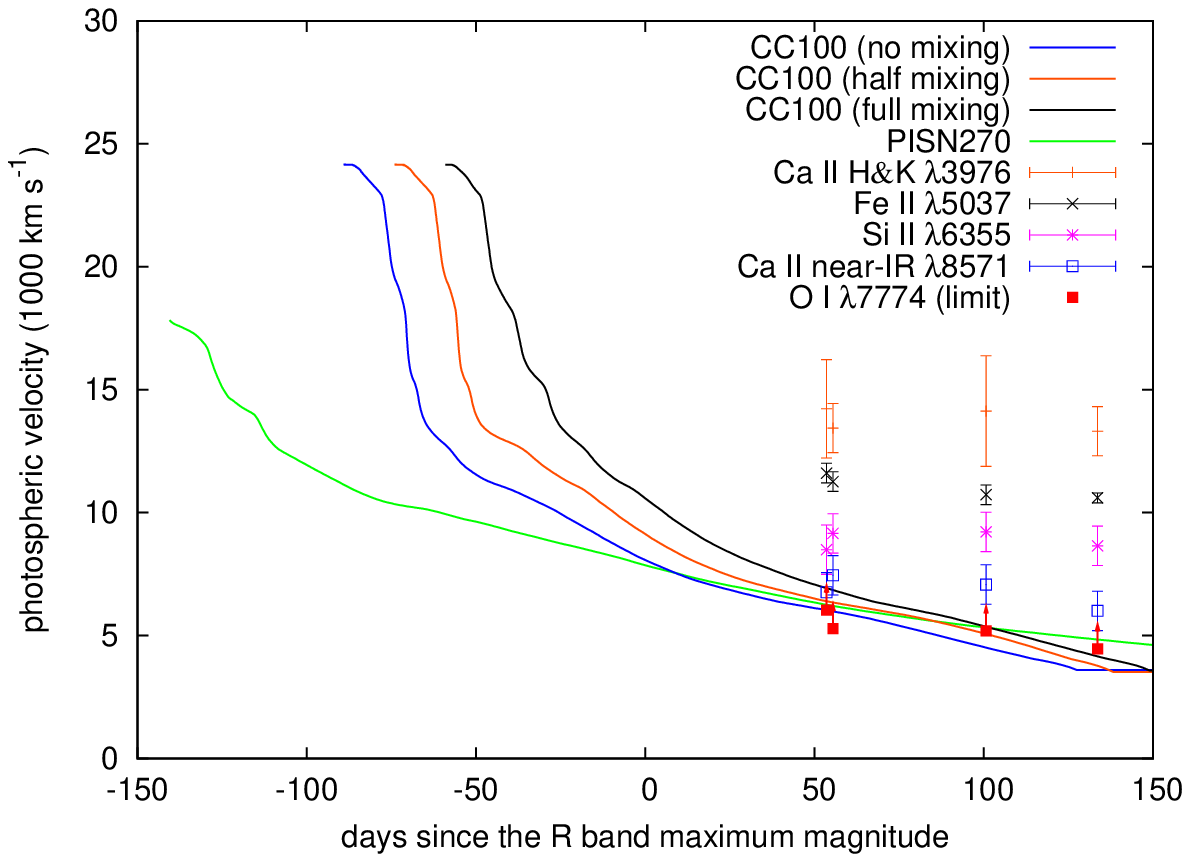}
\caption{
Photospheric velocities of the core-collapse SN and PISN models.
The horizontal axis shows the days in the rest frame.
The line velocities of SN 2007bi observed by Y10
are also shown for comparison.
The line velocities of \ion{O}{i} $\lambda 7774$ shows the lower limit.
All the photospheric velocities are low enough to be
consistent with the observed line velocities.
}\label{07bi-ccPV}
\end{center}
\end{figure}

\section{Core-Collapse Supernova Models for SN 2007bi}
We construct several core-collapse SN models and compare with
the observations of the bolometric LC and the line velocities of SN 2007bi shown
in Y10.
Since the LC of Y10 does not cover the rising part of the LC,
we estimate the bolometric magnitude of the rising part
from the $R$ band observations (G09) assuming the same
bolometric correction (0.45 mag) as in the $R$ band maximum.
We also take into account mixing
since it is possible that a jet emerges from the central remnant
and causes the mixing of the ejecta ({\it e.g.}, Maeda \& Nomoto 2003,
Tominaga 2009).

The LCs of successful models (CC100) are shown in Figure \ref{07bi-ccLC}.
The kinetic energy (\Ekin), ejecta mass (\Mej), and \Ni~mass (\Mni) in
the ejecta are 
\Ekin~=~$3.6\times 10^{52}$ erg, \Mej~=~40~\Msun, and \Mni~=~6.1~\Msun,
which are the same in all the models.
The mass cut between the ejecta and the compact remnant is set at $M_r=3$ \Msun, where $M_r$ is the
mass coordinate, so that
the ejecta contains 6.1~\Msun~of \Ni, which turns out to be
consistent with the bolometric LC of SN 2007bi.
The kinetic energy \Ekin~needs to be large to produce \Mni~=~6.1 \Msun.
The mass of some elements in the ejecta are summarized in
Table \ref{elem}.  The kinetic energy is as large as
those of previously observed SNe which were
associated with a gamma-ray burst
 (SNe 1998bw, 2003dh, 2003lw; {\it e.g.}, Nomoto et al. 2006).

We adopt two different degrees of mixing to see its effects on the LC.
The full-mixing model assumes that the whole ejecta are uniformly mixed.
The half-mixing model assumes that the inner half of the ejecta
(in the mass coordinate) is uniformly mixed.
One of the effects of the mixing is seen in the rise time of the LC.
With mixing, \Ni~is distributed closer to the surface of the ejecta, 
so that the diffusion time is shorter 
and the rise time becomes shorter.
The rise time of the model without mixing is 85 days while
the rise times of the half-mixing model and the full-mixing model are
67 days and 52 days, respectively.
As the rise time of SN 2007bi is not observationally well-determined,
all the models
are consistent with the bolometric LC of SN 2007bi.
The initial decline part of the calculated LCs before maximum 
is formed by the shock heating of the envelope and
its subsequent cooling due to rapid expansion.
Radiation hydrodynamical calculations are required to obtain the realistic
LC at this epoch.

In Figure \ref{07bi-ccPV}, we show the photospheric velocities obtained
by the LC calculations. With the photospheric velocities, we also show
the observed line velocities of SN 2007bi taken from Figure 17 of Y10.
The photospheric velocities of all the models are 
consistent with the observed lowest line velocities, which are thought
to trace the photospheric velocity.

One of the big difference between the core-collapse SN models
and the PISN models is
the abundance of the elements like Si and S.  The abundance of our
core-collapse SN model is consistent with the directly estimated
abundances from the one-zone model of G09 adopted to the
observed emission lines,
{\it i.e.}, C, O, Na, Mg, Ca, and \Ni.
However, it differs significantly in elements
whose abundances
are only indirectly constrained by the spectra (such as Si and S) as
these may have played
a role in line cooling processes. In order to confirm that the abundance
of the core-collapse SN model are consistent with the nebular spectra,
we have to perform spectral synthesis calculations
for the realistic hydrodynamical model of ejecta
rather than the single-zone adopted by G09.
As Si and S have many emission lines in the
infrared range, infrared spectra are also helpful to distinguish PISNe
from core-collapse SNe.
We also point out that, if SN 2007bi is confirmed to be a PISN,
we could expect that PISNe played a role in the chemical enrichment
in the early Universe and there should be some old stars with
chemical compositions expected from PISNe, although they are still not
discovered ({\it e.g.}, Cayrel et al. 2004).

\begin{deluxetable}{ccccccccc}
\tablecaption{The Amount of Elements Contained in the Ejecta}
\tablehead{
$^{12}$C&$^{16}$O&$^{20}$Ne&$^{24}$Mg&$^{28}$Si&$^{32}$S&$^{36}$Ar&$^{40}$Ca&\Ni
}
\startdata
1.4&18.7&1.4&1.5&5.1&2.7&0.5&0.4&6.1
\enddata
\tablecomments{Units: \Msun}
\label{elem}
\end{deluxetable}

\section{Conclusions and Discussion}

In this Letter, we have shown that the LC and photospheric velocity of
SN 2007bi are well-reproduced by the core-collapse SN model
CC100.  
As some gamma-ray bursts are connected to such high energy Type Ic
SNe, the extremely luminous SNe like SN 2007bi
could also be connected to gamma-ray bursts which result from very
massive stars. If this is the case, extremely luminous SNe like SN 2007bi
could be connected to gamma-ray bursts of much more massive star origin
than known SNe associated with a gamma-ray burst.
Even stars more massive than 300 \Msun~could be
the origin of luminous SNe ({\it e.g.}, Ohkubo et al. 2006, 2009).

We note, however, that, although SN 2007bi may not necessarily be a
PISN, the observational data available for SN 2007bi is not sufficient
to single out the explosion mechanism.  In fact, Kasen \& Bildsten
(2009) suggested that the magnetar-powered light curve model (also,
Maeda et al. 2007; Woosley 2009) might explain the LC of SN 2007bi.

Here we show the comparison between our PISN model and SN 2007bi, and
discuss how to distinguish the models for luminous SNe.  We
also apply such LC comparison to SN 2006gy.

\begin{figure}[t]
\begin{center}
\plotone{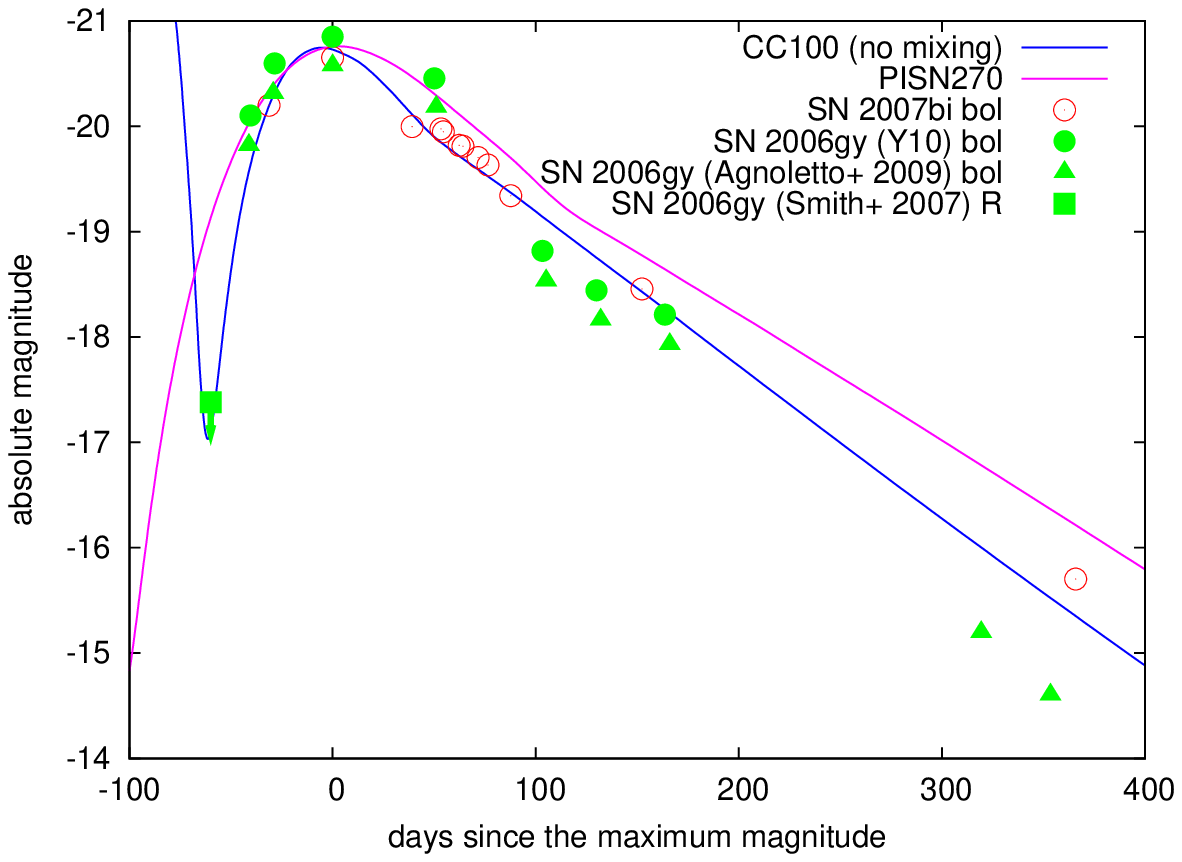}
\caption{
A PISN model for SN 2006gy (PISN270).
The bolometric LC of SN 2006gy is taken from Y10 and Agnoletto et
 al. (2009).
The first point of SN 2006gy is the $R$ band upper limit from Smith et
 al. (2007).
The CC100 no-mixing model and observed bolometric LC of SN 2007bi in Figure
 \ref{07bi-ccLC} are also shown.
The horizontal axis shows the days in the rest frame.
}\label{fig3}
\end{center}
\end{figure}

\subsection{Pair-Instability Supernova Models for SN 2007bi}

In Section 3 (Figure \ref{07bi-ccLC}), we have shown that observations of SN
2007bi are well reproduced by the core-collapse SN model
(CC100).  Here we confirm the claim made by
G09 that a PISN model can also be consistent
with the bolometric LC of SN 2007bi by using the approximate PISN
model PISN270.

The PISN270 model is constructed by scaling the physical structure of
the homologously expanding model CC100 to the ejecta model with
\Mej~=~121 \Msun~and \Ekin~=~$7\times 10^{52}$ erg.  The ejecta mass
\Mej~is the same as the C+O core mass of the PISN model with
\Mms~=~270 \Msun~(UN02), and \Ekin~is obtained from the nuclear energy
released by explosive nuclear burning of the C+O core (UN02).  Here
the same amount of \Ni~(\Mni~=~9.8~\Msun) as in the 270 \Msun~ model
(UN02) is assumed to be synthesized in the inner layers.  Note that
the 270 \Msun~ model of UN02 still has the H-rich and He envelopes at
the time of explosion and, here, we assume that the envelopes were
stripped off by some mechanism.

Figure \ref{fig3} shows that the bolometric LC of PISN270 (the red line)
is consistent with the bolometric LC of SN 2007bi (red open circles).
The rise time to the LC peak for PISN270 is $\sim 150$ days, being
consistent with the PISN model in G09.  This rise time is longer than
the core-collapse SN model CC100 (Figure \ref{fig3}), because
the photon diffusion takes more time in more massive PISN270.
Although \Mni~of PISN270 is $\sim$ 1.6 times larger than that of the
core-collapse SN CC100 model, the longer rise time lowers the peak
brightness powered by the radioactive decay.  These two effects make
the peak magnitude of PISN270 similar to CC100.

This difference in the rising part of the LC is important to
discriminate between the core-collapse SN and the PISN models.
Although SN 2007bi was not observed early enough, much earlier
observations before the peak could constrain the SN type from the LC.
In addition, as already mentioned in Section 3,
the abundance of Si and S would also be a key to
distinguish between the two models.

\subsection {Models for SN 2006gy}

As mentioned in Section 1, there has been some suggestions that the luminous
Type IIn SN 2006gy is a PISN ({\it e.g.}, Smith et al. 2007).  We
thus apply our LC models for comparison with SN 2006gy.  Figure
\ref{fig3} shows the bolometric LC of SN 2006gy (filled circles and
triangles).  As only the $R$ band magnitude was observed in the early
epochs of SN 2006gy (the filled square), we cannot construct the
bolometric LC at the early epochs but we can constrain the rise time
of the LC.  Our calculations show that the rise time of the PISN model
is too slow to be consistent with SN 2006gy.  Although our PISN model
does not have a H-rich envelope, the presence of the H-rich envelope
could even slow down the brightening ({\it e.g.}, Kawabata et al. 2009).
Woosley et al. (2008) showed that the
interaction between the pulsating core and the envelope can power the
LC of SN 2006gy.  As a similar mechanism, the interaction of a SN
ejecta with its very dense circumstellar matter could convert the
kinetic energy of ejecta directly to radiation energy and
could also be the origin a luminous SN like SN 2006gy.

\begin{acknowledgments} 
We thank the anonimous referees for their advice which improved the text.
Numerical calculations were carried out on the general-purpose PC farm
at Center for Computational Astrophysics, CfCA, of National
Astronomical Observatory of Japan.
This research has been supported in part by World Premier
International Research Center Initiative, MEXT, and by the
Grant-in-Aid for Scientific Research of the JSPS (18104003, 20540226, 20840007)
and MEXT (19047004, 22012003), Japan.
\end{acknowledgments}


\bibliographystyle{apj}

\end{document}